\documentclass[11pt,twoside]{article}
\usepackage{amsmath,amsfonts,amssymb}
\usepackage{latexsym}
\usepackage{dcolumn}
\usepackage{graphicx,epsfig}
\usepackage{amsthm}
\usepackage{color}

\begin{document}

\setcounter{page}{1}

\pagestyle{plain} \vspace{1 cm}

\begin{center}
{\large{\bf {A Quantum-Corrected Approach to the Black Hole Radiation via a Tunneling Process}}}\\

\vspace{1 cm}
{\bf Milad Hajebrahimi$^{\dag,}$}\footnote{m.hajebrahimi@stu.umz.ac.ir}\quad and \quad {\bf Kourosh Nozari$^{\dag,}$}\footnote{knozari@umz.ac.ir}\\
\vspace{0.5 cm}
$^{\dag}$Department of Theoretical Physics, Faculty of Basic Sciences, University of Mazandaran,\\
P. O. Box 47416-95447, Babolsar, Iran\\

\end{center}

\vspace{1.5cm}

\begin{abstract}
In the language of black hole physics, Hawking radiation is one of the most controversial subjects about which there exist lots of puzzles, including the information loss problem and the question of whether this radiation is thermal or not. In this situation, a possible way to face these problems is to bring quantum effects into play, also taking into account self-gravitational effects in the scenario. We consider a quantum-corrected form of the Schwarzschild black hole inspired by the pioneering work of Kazakov and Solodukhin to modify the famous Parikh–Wilczek tunneling process for Hawking radiation. We prove that in this framework the radiation is not thermal, with a correlation function more effective than the Parikh–Wilczek result, and the information loss problem can be addressed more successfully. Also, we realize that quantum correction affects things in the same way as an electric charge. So, it seems that quantum correction in this framework has something to do with the electric charge. 
\end{abstract}

\newpage
\section{Introduction}

Since the prediction of the existence of black holes through the general theory of relativity proposed by Albert Einstein in 1915, these amazing and mysterious objects have became one of the most controversial research areas of fundamental physics. A vast body of research work has focused on the creation, formation, characteristics, possibility of existence, and even detection of these black, and at the same time apparently radiating, holes in the universe \cite{Hawking1972, Bekenstein1973, Hawking1974, Hawking1976, Chandrasekhar1985, Kormendy1995}. Understanding these subjects about black holes may help us to understand the creation and beginning of the universe, i.e. the very early epoch of the universe where quantum gravity is the dominant theory on the creation and evolution of the cosmos \cite{Carr1974}. Recently, following the first ever image of a supermassive black hole in the center of the M87 galaxy by the Event Horizon Telescope team \cite{1EHTC2019, 2EHTC2019, 3EHTC2019, 4EHTC2019, 5EHTC2019, 6EHTC2019}, people have come to believe in their existence and hope to comprehend the universe through understanding these surprising holes.

One of the main subjects concerning black holes is the possibility of radiating energy into space so that a far observer can detect it. Such radiation is completely thermal, with temperature $$T_{H}=\frac{1}{8\pi M}$$ known as the Hawking temperature, in which M is the black hole mass. Due to the conservation law of energy, the black hole mass has to decrease as it radiates energy into space. This mass decrease is known as “Hawking evaporation.” Such evaporation continues until $M=0$ when the whole mass of the black hole has been radiated. In this situation, all of the information within the black hole will be lost through this thermal radiation, which leads to the “information loss problem.” This idea was first proposed fully by Hawking in 1975 \cite{Hawking1975}, when the origin of this radiation (known as Hawking radiation) was not specified. So, lots of issues were proposed in the literature of fundamental physics to address these puzzles.

It is thought that these puzzles will be addressed in a theory of quantum gravity which has not yet been proposed. But, some proposals towards this theory are now available \cite{Amelino2013}. On the other hand, some arguments that insert the outcomes of quantum mechanics into the framework of the general theory of relativity can address the aforementioned puzzles. One of these expresses the Hawking radiation as a tunneling process in a direct way. This process is known as Parikh–Wilczek tunneling, in which, unlike what Hawking suggested, particles are in a dynamical geometry to implement the energy conservation \cite{Parikh2000}. The term dynamical geometry indicates the black hole mass, and consequently its radius is going to shrink as the black hole radiates energy into space. By respecting conservation laws, this proposal has managed to address the information loss problem to some extent, and it has been shown that, through the tunneling process, Hawking radiation is no longer thermal.

All types of black holes have a central point in their spherical geometry, which is known as the intrinsic singularity point, and, through the cosmic censorship hypothesis \cite{Penrose1969}, one or more apparently singular boundaries called event horizons from which nothing, even light, can escape, as was first introduced by Karl Schwarzschild in 1916. The singularity point is not physically meaningful, so a lot of work has focused on proposing a way to deform the Schwarzschild black hole to produce a regular black hole. One way is to insert the outcomes of quantum field theories into the framework, as Kazakov and Solodukhin did in 1994 \cite{Kazakov1994}. By taking into account spherically symmetric quantum fluctuations (excitations) of the background metric and a typical matter field, they modified the Schwarzschild black hole so that there is no Ricci flatness outside of it and also the intrinsic singularity point at $r=0$ is smeared over a two-sphere with a radius of $r\sim r_{pl}$, which leads to a regular black hole.

In this study, we consider a quantum-corrected Schwarzschild black hole in the spirit of the pioneering work of Kazakov and Solodukhin. Then, we reconsider the Parikh–Wilczek tunneling process in detail to determine the role of quantum effects arising from quantum field theories on this process. Since such a black hole has two event horizons, we implement the method used in Ref. \cite{Yan-Gang2011} for expanding the metric coefficient to deform the contour in the contour integrations in calculating the emission rate.

We start the procedure by introducing the quantum-corrected Schwarzschild black hole by writing its line element. Then we first proceed for massless neutral particles which move on null geodesics and then for massive neutral particles traveling on time-like world-lines. Since the particles and the hole are neutral, there is no electrical interaction between particles and the black hole at all. The quantum correction incorporated in this manner seems to do something with the electric charge since it resembles a metric similar to the Reissner–Nordstr\"{o}m metric. Finally, we end with some conclusions.

\section{Quantum-corrected Schwarzschild black hole}

In recent decades physicists did not think it plausible that a singularity in spacetime could exist at the center of a black hole, as in the Schwarzschild solution. Among the many efforts to remove this intrinsic singularity, one of them proceeded by taking into account spherically symmetric quantum fluctuations of the background metric, introducing a quantity ``$a$'' which brings a quantum correction term into the solution. This quantity has the dimensionality of length that leads to a quantum, spherically symmetric deformation of the Schwarzschild solution as \cite{Kazakov1994}
\begin{equation}
ds^{2}=-g_{_{00}}dt^{2}+\frac{dr^{2}}{g_{_{00}}}+r^2d\Omega^{2},
\end{equation}
where
\begin{equation}
g_{_{00}}=-\frac{2M}{r}+\frac{1}{r}\sqrt{r^{2}-a^{2}}
\end{equation}
and $M$ is the mass of the black hole. While the singularity point in the Schwarzschild black hole was at $r=0$, now it is shifted to the finite radius $r=a$ which means that instead of a ``point'' that singularity is now ``spread'' over a two-dimensional sphere of radius $r=a$.
By asymptotic expression of Eq. (2) for large $r$ (i. e. $r\gg a$) and eliminating higher-order terms of $a$, one gets a simplified form
\begin{equation}
g_{_{00}}\approx1-\frac{2M}{r}-\frac{a^{2}}{2r^{2}}.
\end{equation}
One can easily conclude that the black hole has two event horizons associated with the metric coefficient in Eq. (3), an inner horizon and an outer horizon:
\begin{equation}
r_{inner}=\frac{1}{2}\Big[2M-\sqrt{4M^2+2a^2}\Big]
\end{equation}
\begin{equation}
r_{outer}=\frac{1}{2}\Big[2M+\sqrt{4M^2+2a^2}\Big]\,,
\end{equation}
resembling a Reissner-Nordstr\"{o}m electrically charged black hole. In 1999, Parikh and Wilczek propounded a direct explanation for Hawking radiation as a tunneling process of particles through the hole’s horizon in a ``dynamical geometry" \cite{Parikh2000}. They precisely showed that the barrier associated with this tunneling is the energy of the particle itself. They also proved that due to the tunneling, the radius of the black hole decreases, hence the geometry is dynamic. This idea has been generalized to address the information loss problem by incorporating phenomenological quantum gravity effects via a generalized uncertainty principle \cite{Nozari2008a} and also non-commutative geometry \cite{Nozari2008b}.

Since the tunneling process will occur across the horizon, we have to regularize the coordinates at this one-way boundary. The Schwarzschild coordinates are singular at the horizon, so with respect to the expanded metric coefficient in Eq. (3) one can define a new time coordinate as
\begin{equation}
t_{p}=t+\int\bigg[\frac{\sqrt{1-g_{_{00}}}}{g_{_{00}}}\bigg]dr,
\end{equation}
which leads to the Painlev\'{e} coordinates \cite{Painleve1921}, a suitable coordinates system in which the metric is well-defined, dynamic and stationary. By inserting Eq. (6) into Eq. (1) and then equating the coefficient of $dr^{2}$ to unity, one can get the quantum-corrected Schwarzschild line element in the Painlev\'{e} coordinates as
\begin{equation}
ds^{2}=-\bigg[1-\frac{2M}{r}-\frac{a^{2}}{2r^{2}}\bigg]dt_{p}^{2}+2\bigg[\sqrt{\frac{2M}{r}+\frac{a^{2}}{2r^{2}}}\bigg]drdt_{p}
+dr^{2}+r^{2}d\Omega^{2}.
\end{equation}
In what follows we proceed with two different cases: the first case considers massless virtual particles which move on null geodesics at light velocity, and the second case is devoted to massive virtual particles which move on time-like geodesics at a velocity smaller than the light velocity. In both cases, the virtual particles are considered as spherically symmetric shells with energy $\omega$.

\section{Tunneling process for massless particles}

One can deduce the equation of radial motion of a massless particle with respect to the line element in Eq. (7) as
\begin{equation}
\dot{r}\equiv\pm1-\sqrt{\frac{2M}{r}+\frac{a^{2}}{2r^{2}}}
\end{equation}
in which the plus sign is for outgoing motion, the minus sign for ingoing motion and the dot stands for differentiation with respect to the proper time, $\tau=t_{p}$.  We consider that the pair creation of virtual particle-antiparticle would happen just inside the outer event horizon of the hole. The virtual antiparticle with negative energy is absorbed in the hole, so the hole becomes smaller in size. The massless virtual particle with positive energy will tunnel out of the hole, outside the outer horizon. For this, the outgoing null geodesics should be considered. In this situation, the particle's self-gravitation effects could be added. In this study, we leave the hole's mass to fluctuate, so instead of $M$, we must substitute $(M-\omega)$ in the following equations because of the decrease in the hole's mass during the tunneling by emission of a virtual particle with energy $\omega$. In this manner, the Eqs. (7) and (8) with self-gravitating effects take the following forms respectively:
\begin{equation}
ds^{2}=-\bigg[1-\frac{2(M-\omega)}{r}-\frac{a^{2}}{2r^{2}}\bigg]dt_{p}^{2}+2\bigg[\sqrt{\frac{2(M-\omega)}{r}
+\frac{a^{2}}{2r^{2}}}\bigg]drdt_{p}+dr^{2}+r^{2}d\Omega^{2}
\end{equation}
\begin{equation}
\dot{r}\equiv1-\sqrt{\frac{2(M-\omega)}{r}+\frac{a^{2}}{2r^{2}}},
\end{equation}
where $\omega$, as mentioned above, is the energy of the virtual particle that is taken to be the same as the potential barrier in the tunneling process.

One can consider that virtual particle is initially at, $r_{in}$, which is an $\epsilon$ smaller than the outer event horizon radius, and then crosses it to somewhere, $r_{out}$, which is an $\epsilon$ greater than the reduced outer horizon radius ($\epsilon$ is a very small quantity with dimension of length)
\begin{equation}
r_{in}=\frac{1}{2}\Big[2M+\sqrt{4M^2+2a^2}\Big]-\epsilon
\end{equation}
\begin{equation}
r_{out}=\frac{1}{2}\Big[2(M-\omega)+\sqrt{4(M-\omega)^2+2a^2}\Big]+\epsilon.
\end{equation}
At the outer horizon, the radial wavenumber of the virtual particle becomes infinite and its wavelength is extremely blue-shifted. Consequently, the WKB (Wentzel-Kramers-Brillouin) approximation is valid. So, in this semiclassical procedure, as explained above, one can find the transmission coefficient (emission rate) of the tunneling (radiating) particles approximately as an exponential function of the imaginary part of the action
\begin{equation}
\Gamma\approx e^{-2\, \operatorname{Im} S}.
\end{equation}
The imaginary part of the action for this virtual particle can be expressed (see the details in Ref. \cite{Parikh2000}) as
\begin{equation}
\operatorname{Im} S =\operatorname{Im} \int_{r_{in}}^{r_{out}}p_{r}dr=\operatorname{Im} \int_{r_{in}}^{r_{out}}\int_{0}^{p_{r}}d\tilde{p}_{r}dr\,,
\end{equation}
where $p_{r}$ is the canonical momentum of the virtual particle. By using Hamilton's equation
\begin{equation}
\dot{r}=\frac{dH}{dp_{r}}\,,
\end{equation}
where $H=M-\tilde{\omega}$ is the Hamiltonian, one can rewrite Eq. (14) as
\begin{equation}
\operatorname{Im} S =\operatorname{Im} \int_{0}^{\omega}\int_{r_{in}}^{r_{out}}\frac{dr(-d\tilde{\omega})}{\dot{r}}\,.
\end{equation}
Inserting Eq. (10) into Eq. (16), one can see that the integrand has two poles and it should be solved by deforming the contour. For simplicity, one can expand $g_{_{00}}$ around the outer horizon $r_{outer}$, to make it have just one pole that is located at $r_{outer}$ \cite{Yan-Gang2011}. Therefore, the above integral could be easily solved. The expanded $g_{_{00}}$ is of the form
\begin{equation}
g_{_{00}}=g_{_{00}}(r)\approx g_{_{00}}(r_{outer})+g'_{_{00}}(r_{outer})(r-r_{outer})
\end{equation}
where
\begin{equation}
g'_{_{00}}(r_{outer})\equiv\frac{dg_{_{00}}}{dr}\Big |_{r=r_{outer}}=\frac{r_{outer}-r_{inner}}{r_{outer}^{2}}
\end{equation}
Finally, the result of the radial integral can be expressed as
\begin{equation}
\operatorname{Im} S =\int_{0}^{\omega}\frac{2\pi\Big[2(M-\tilde{\omega})+\sqrt{4(M-\tilde{\omega})^{2}}\Big]^{3}}
{4a^{2}+4(M-\tilde{\omega})\Big[2(M-\tilde{\omega})+\sqrt{4(M-\tilde{\omega})^{2}}\Big]}d\tilde{\omega}.
\end{equation}
Next, this integral can be performed easily to achieve the following final result
\begin{equation}
\operatorname{Im} S=4\pi\omega\Big(M-\frac{\omega}{2}\Big)-\pi\big(M-\omega\big)\sqrt{4\big(M-\omega\big)^{2}+2a^{2}}
+\pi M\sqrt{4M^{2}+2a^{2}}
\end{equation}
The same result could be derived for ingoing radial motion of a virtual particle with negative energy which is created outside the black hole and tunnels into it. Now, by applying Eq. (20) in Eq. (13), one can find the emission rate as
\begin{eqnarray}
\Gamma\approx
\exp\bigg\{-4\pi\bigg[2\omega(M-\frac{\omega}{2})-\frac{1}{2}(M-\omega)\sqrt{4(M-\omega)^{2}+2a^{2}}
+\frac{1}{2}M\sqrt{4M^{2}+2a^{2}}\bigg]\bigg\}=e^{+\Delta S_{BH}}
\end{eqnarray}
where $\Delta S_{BH}$ is the change in the Bekenstein-Hawking entropy due to particle emission. Figure 1 shows the variation of $\ln(\Gamma)$ versus $M$ in which the line of the quantum-deformed case goes to negative values faster than the ordinary case.
\begin{figure}
\begin{center}\includegraphics{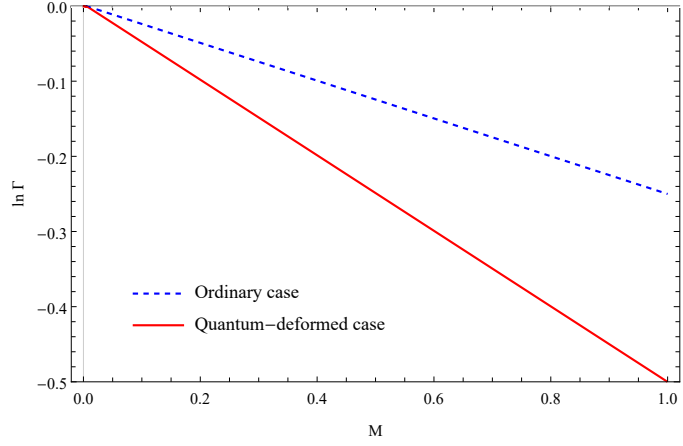} \vspace{7 cm}
\end{center}
\caption{\label{Fig1}\small {Logarithm of emission rate versus the black hole mass. The blue dashed line depicts the ordinary Parikh-Wilczek result and the red solid line presents the quantum-deformed case established in our study. In drawing this figure we have set $\omega=0.01$ and $a=0.0001$ in order to have a qualitative intuition.}}
\end{figure}

Conceivably, this result is completely equivalent to the tunneling process for a Reissner-Nordstr\"{o}m black hole as derived in Ref. \cite{Parikh2000} by letting $$Q\equiv i\frac{\sqrt{2}}{2}a\,,$$ where $Q$ is the charge parameter in the Reissner-Nordstr\"{o}m black hole. This analogy tempts us to conclude that the electric charge behaves in the same way as quantum effect in this setup. This is an interesting result and has also been observed in \cite{Nozari2008c} in a non-commutative background. There, the authors showed that thermodynamically a \emph{non-commutative} Schwarzschild black hole
behaves in much the same way as a \emph{commutative} Reissner-Nordstr\"{o}m black hole. So, it seems that electric charge has something to do with the very structure of the spacetime manifold and shows itself as a quantum gravitational effect. It seems that the existence of a point charge in spacetime structure leads to quantum fluctuations of the background manifold.

One can compare the $e^{-2\,\operatorname{Im} S}$ with the Boltzmann factor $e^{-\beta\omega}$ by dropping the higher-order terms of $\omega$ in Eq. (21) for large $M$, that is, $M\gg \omega$, to find a modified Hawking Temperature:
\begin{equation}
T_{H,massless}=\frac{1}{2\pi}\frac{\sqrt{4M^{2}+2a^{2}}}{\big(2M+\sqrt{4M^{2}+2a^{2}}\big)^{2}}.
\end{equation}
As stated above, this is specifically in agreement with the outcome of the deformed Hawking Temperature emerging from the tunneling process for a Reissner-Nordstr\"{o}m charged black hole with $Q=i\frac{\sqrt{2}}{2}a$.

In Fig. 2, the behavior of $T$ versus $M$ is shown for the non-deformed and deformed cases. Our rigorous inspection (by numerical treatment) shows that the ordinary Hawking evaporation terminates with divergent temperature without any remnant. However, the quantum-deformed Hawking evaporation (with $a>1$) terminates with a non-zero mass remnant with a finite temperature. In fact, as a minimal length parameter in this scenario, the radius $r=a$ is the smallest radius of this quantum-deformed, non-singular Schwarzschild black hole. Hence, this black hole should radiate the Hawking radiation until it approaches the parameter $a$. Consequently, the quantum deformation stops the black hole from complete evaporation (see for instance Ref. \cite{Chen2015} for a similar situation in the generalized uncertainty principle). Also, unlike what was deduced by Parikh and Wilczek, in this study it seems that without eliminating higher-order terms of $\omega$, the non-thermality of Hawking radiation can be addressed. It is worth noting that the choice of $a>1$ is just for simplicity in plotting Fig. 2 qualitatively and clearly showing the remnant.
\begin{figure}
\begin{center}\includegraphics{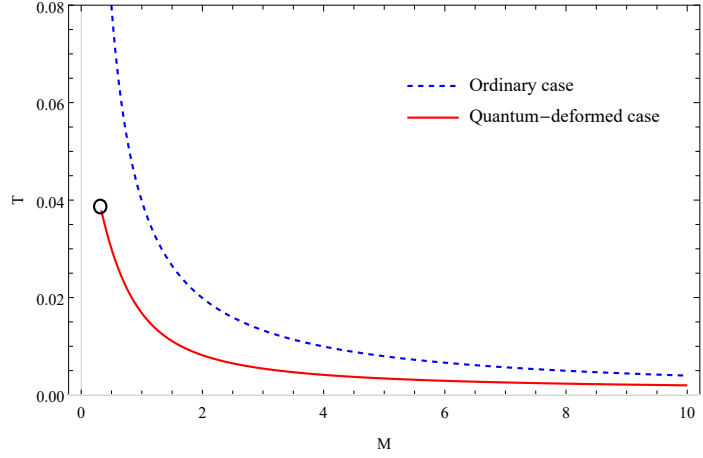} \vspace{7 cm}
\end{center}
\caption{\label{Fig2}\small {Black hole temperature versus its mass. The blue dashed line depicts the ordinary Parikh-Wilczek result and the red solid line presents the quantum-deformed case established in our study. We have set $a=2$ to provide a qualitative intuition.}}
\end{figure}

One can see that the black hole's information loss problem can be addressed in this scenario due to the presence of quantum effects. To prove it, the correlation function between emitted particles (modes) should be determined. In this case, the correlation function could be expressed as
\begin{equation}
\chi(\omega_{1}+\omega_{2};\omega_{1},\omega_{2})\equiv\ln[\Gamma(\omega_{1}+\omega_{2})]-\ln[\Gamma(\omega_{1})\Gamma(\omega_{2})]
\end{equation}
where $\omega_{1}$ and $\omega_{2}$ are the energies of the particles labeled ``1'' and ``2'', respectively. This leads simply to the following expression for the correlation function between the emitted particles:
\begin{eqnarray}
\chi(\omega_{1}+\omega_{2};\omega_{1},\omega_{2})=2\pi M\sqrt{4M^{2}+2a^{2}}+4\pi\bigg\{\Big[\Big(2M-\omega_{1}\Big)\omega_{1}-\frac{1}{2}\Big(M-\omega_{1}\Big)
\sqrt{4\big(M-\omega_{1}\big)^{2}+2a^{2}}\Big]\nonumber\\
+\Big[\Big(2M-\omega_{2}\Big)\omega_{2}-\frac{1}{2}\Big(M-\omega_{2}\Big)\sqrt{4\big(M-\omega_{2}\big)^{2}+2a^{2}}\Big]
\nonumber\\-\Big[\Big(2M-(\omega_{1}+\omega_{2})\Big)\Big(\omega_{1}+\omega_{2}\Big)
-\frac{1}{2}\Big(M-(\omega_{1}+\omega_{2})\Big)\sqrt{4\Big(M-(\omega_{1}+\omega_{2})\Big)^{2}+2a^{2}}\Big]\bigg\}
\end{eqnarray}
which is obviously not zero and therefore radiation is not completely thermal. This means that the tunneling probability of these two particles with energies $\omega_{1}$ and $\omega_{2}$ is not the same as tunneling probability of a particle with total energy $(\omega_{1}+\omega_{2})$. Consequently, there is a correlation between radiant quanta, and thus one could assemble the information radiating from the black hole. In other words, part of information is assembled in the correlation between emitted particles. This is possibly a promising way of addressing the information loss problem. Also, with respect to the correlation function $$\chi=8\pi\omega_{1}\omega_{2}$$ deduced from Parikh and Wilczek’s work, in this study the role of the quantum deformation is to strengthen correlation function between radiating massless particles.

\section{Tunneling Process for Massive Particles}

For a massive particle, instead of the null trajectories one has to determine time-like trajectories by using the particle's Lagrangian. In this case, since the particle does not have an electric charge itself, its trajectory is truly geodesic. The scheme for deriving such a geodesic is definitely obtained in Ref. \cite{Yan-Gang2011}. Following this scheme, one can write the Lagrangian of the particle (which is actually the kinematic energy of the particle) as
\begin{equation}
2K=mg_{_{\mu\nu}}\frac{dx^{\mu}}{d\tau}\frac{dx^{\nu}}{d\tau}
\end{equation}
where $m$ is the particle's mass and $\tau$ is the proper time. Accordingly, for the radial motion of the particle with respect to the line element in Eq. (7), the Lagrangian in Eq. (25) would take the following form:
\begin{equation}
2K=m\bigg(-\Big[1-\frac{2M}{r}-\frac{a^{2}}{2r^{2}}\Big]\dot{t}^{2}_{p}+2\Big[\sqrt{\frac{2M}{r}+\frac{a^{2}}{2r^{2}}}\Big]
\dot{t}_{p}\dot{r}+\dot{r}^{2}\bigg).
\end{equation}

On the other hand, the time-like interval between two points in the spacetime introduced by the line element in Eq. (7) is of the form $g_{_{\mu\nu}}\dot{x}^{\mu}\dot{x}^{\nu}=-1$. Consequently, one can achieve the following two equations:
\begin{equation}
P_{t_{p}}\equiv-\frac{\partial K}{\partial \dot{t}}=m\Big[1-\frac{2M}{r}-\frac{a^{2}}{2r^{2}}\Big]\dot{t}_{p}-
m\Big[\sqrt{\frac{2M}{r}+\frac{a^{2}}{2r^{2}}}\Big]\dot{r}=cte.\equiv\omega\,,
\end{equation}
\begin{equation}
-\Big[1-\frac{2M}{r}-\frac{a^{2}}{2r^{2}}\Big]\dot{t}^{2}_{p}+2\Big[\sqrt{\frac{2M}{r}+\frac{a^{2}}{2r^{2}}}\Big]
\dot{t}_{p}\dot{r}+\dot{r}^{2}=-1\,,
\end{equation}
where $P_{t_{p}}$ is the particle's momentum in the Painlev\'{e} time coordinate. It is simple to solve this set of equations to find two unknowns, i. e. $\dot{t}_{p}$ and $\dot{r}$ as follows:
\begin{equation}
\dot{r}=\pm\frac{1}{m}\sqrt{\omega^{2}-m^{2}\Big[1-\frac{2M}{r}-\frac{a^{2}}{2r^{2}}\Big]}
\end{equation}
\begin{equation}
\dot{t}_{p}=\frac{1}{m\Big(1-\frac{2M}{r}-\frac{a^{2}}{2r^{2}}\Big)}\bigg[\omega+\sqrt{\Big(\frac{2M}{r}
+\frac{a^{2}}{2r^{2}}\Big)\Big(\omega^{2}-m^{2}\Big[1-\frac{2M}{r}-\frac{a^{2}}{2r^{2}}\Big]\Big)}\bigg].
\end{equation}

Since we suppose that the particle-antiparticle pair is created inside the outer horizon, as mentioned in the previous section, the plus sign of Eq. (29) is taken into account to represent the outward motion of the particle. Therefore, the radial outward time-like geodesic for a particle with mass $m$ and energy (or momentum) $\omega$ in such a spacetime in the Painlev\'{e} coordinates could be expressed as
\begin{equation}
\frac{dr}{dt_{p}}\equiv\frac{\dot{r}}{\dot{t}_{p}}=\Big(1-\frac{2M}{r}-\frac{a^{2}}{2r^{2}}\Big)
\frac{\sqrt{\omega^{2}-m^{2}\Big[1-\frac{2M}{r}-\frac{a^{2}}{2r^{2}}\Big]}}{\bigg[\omega+\sqrt{\Big(\frac{2M}{r}
+\frac{a^{2}}{2r^{2}}\Big)\Big(\omega^{2}-m^{2}\Big[1-\frac{2M}{r}-\frac{a^{2}}{2r^{2}}\Big]\Big)}\bigg]}\,.
\end{equation}

Now, one has to introduce self-gravitational effects in the scenario by allowing the mass of the black hole to fluctuate. Then Eq. (31) takes the form
\begin{equation}
\frac{\dot{r}}{\dot{t}_{p}}=\Big(1-\frac{2(M-\omega)}{r}-\frac{a^{2}}{2r^{2}}\Big)
\frac{\sqrt{\omega^{2}-m^{2}\Big[1-\frac{2(M-\omega)}{r}-\frac{a^{2}}{2r^{2}}\Big]}}{\bigg[\omega+
\sqrt{\Big(\frac{2(M-\omega)}{r}+\frac{a^{2}}{2r^{2}}\Big)\Big(\omega^{2}-m^{2}\Big[1-\frac{2(M-\omega)}{r}-
\frac{a^{2}}{2r^{2}}\Big]\Big)}\bigg]}.
\end{equation}
As in the previous case, Eqs. (11) and (12) show the radial positions where, respectively, the massive particle begins to tunnel initially from the inside of the hole, $r_{in}$, and then completes it finally at the outside of the hole, $r_{out}$.

The imaginary part of the action addressed in Eq. (16), is required to calculate the emission rate in Eq. (13). But in this case, Eq. (16) takes the following form:
\begin{equation}
\operatorname{Im} S = \operatorname{Im} \int_{m}^{\omega}\int_{r_{in}}^{r_{out}}\frac{dr(-d\tilde{\omega})}{\Big(\frac{dr}{dt_{p}}\Big)}.
\end{equation}
So, one can apply Eq. (32) in Eq. (33) to gain the imaginary part of the action as
\begin{equation}
\operatorname{Im} S = \operatorname{Im} \int_{m}^{\omega}\int_{r_{in}}^{r_{out}}dr(-d\tilde{\omega})\frac{\bigg[\omega+\sqrt{\Big(\frac{2(M-\omega)}{r}
+\frac{a^{2}}{2r^{2}}\Big)\Big(\omega^{2}-m^{2}\Big[1-\frac{2(M-\omega)}{r}-\frac{a^{2}}{2r^{2}}\Big]\Big)}\bigg]}
{\Big(1-\frac{2(M-\omega)}{r}-\frac{a^{2}}{2r^{2}}\Big)\sqrt{\omega^{2}-m^{2}\Big[1-\frac{2(M-\omega)}{r}-\frac{a^{2}}
{2r^{2}}\Big]}}.
\end{equation}
Again, similar to the previous case, the integrand in Eq. (34) has two poles and by using Eq. (17), one can simplify it to have a pole at the outer horizon. Hence, the radial integral by applying Eq. (18) leads to
\begin{equation}
\operatorname{Im} S=\pi\int_{m}^{\omega}\frac{2}{g'_{_{00}}(r_{outer})}d\tilde{\omega}=
\pi\int_{m}^{\omega}2\frac{r_{outer}^{2}}{r_{outer}-r_{inner}}d\tilde{\omega}
\end{equation}
Finally, the above integral over $\tilde{\omega}$ can be performed to find
\begin{equation}
\operatorname{Im} S=\frac{\pi}{2}\Big[2(M-m)^{2}-2(M-\omega)^{2}+(M-m)\sqrt{4(M-m)^{2}+2a^{2}}-(M-\omega)\sqrt{4(M-\omega)^{2}+2a^{2}}\Big].
\end{equation}
Now the emission rate, by applying Eq. (36) in Eq. (13), is obtained as
\begin{eqnarray}
\Gamma\approx\exp\bigg\{-\pi\Big[2(M-m)^{2}-2(M-\omega)^{2}+(M-m)\sqrt{4(M-m)^{2}+2a^{2}}\nonumber\\
-(M-\omega)\sqrt{4(M-\omega)^{2}+2a^{2}}\Big]\bigg\}=e^{+\Delta S_{BH}}.
\end{eqnarray}
Figure 3 depicts the behavior of $\ln\Gamma$ versus $M$ for both the ordinary case with $a=0$ and the quantum-deformed case. In the ordinary case the black hole evaporates completely due to Hawking Radiation. In the quantum-deformed case the final stage is a non-zero mass remnant. This is actually the case in other approaches such as the generalized uncertainty principle and also non-commutative geometry (see, for instance, Refs. \cite{Nozari2008a,Nozari2008b}).
\begin{figure}
\begin{center}\includegraphics{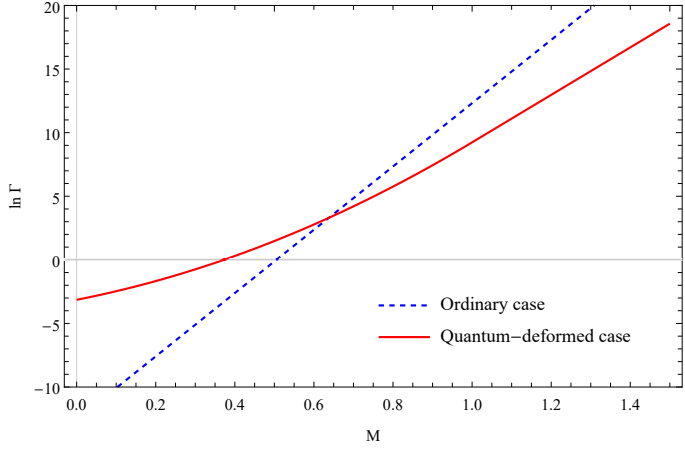} \vspace{7 cm}
\end{center}
\caption{\label{Fig3}\small {The behavior of $\ln\Gamma$ versus $M$. The blue dashed line shows the ordinary case with $a=0$ and the red solid line represents the quantum-deformed case. We have valued $a=0.0001$, $m=1$, and $\omega=0.01$.}}
\end{figure}

Again, one can find the Hawking temperature. By expanding the emission rate in $m$ and $\omega$ to the second order for large $M$, one can deduce the following emission rate:
\begin{eqnarray}
\Gamma\approx\exp\bigg\{\frac{\pi(m-\omega)\big(2M+\sqrt{4M^{2}+2a^{2}}\big)^{2}}{\sqrt{4M^{2}+2a^{2}}}\bigg[1-
\frac{(m-\omega)\big(2M+\sqrt{4M^{2}+2a^{2}}\big)}{4(2M^{2}+a^{2})}-\nonumber\\
\frac{3a^{2}(m-\omega)}{2(2M^{2}+a^{2})\big(2M+\sqrt{4M^{2}+2a^{2}}\big)}\bigg]\bigg\}.
\end{eqnarray}
By comparing the Boltzmann factor with the expanded emission rate for large $M$ and then taking into account the first term of the Eq. (38) with respect to $\omega$, the modified Hawking temperature can be expressed as
\begin{equation}
T_{H,massive}\approx\frac{1}{2\pi}\frac{2\sqrt{4M^{2}+2a^{2}}}{\big(2M+\sqrt{4M^{2}+2a^{2}}\big)^{2}}\,.
\end{equation}
Up to the order of approximation used to derive this relation, $T_{H,massive}$ and $T_{H,massless}$ are the same. Accordingly, the behavior of $T_{H,massive}$ versus $M$ is more or less the same as Fig. 2. Once again, there is a non-zero mass remnant with finite temperature at the final stage of evaporation for the quantum-deformed case, as was expected.

As well as the previous massless case, one can also conclude that for the massive case the Hawking radiation is not thermal. To see this, the correlation function in Eq. (23) can be calculated as
\begin{eqnarray}
\chi=\pi\bigg\{2(M-m_{1})^{2}-2(M-\omega_{1})^{2}+(M-m_{1})\sqrt{4(M-m_{1})^{2}+2a^{2}}
-(M-\omega_{1})\sqrt{4(M-\omega_{1})^{2}+2a^{2}}\nonumber\\
+2(M-m_{2})^{2}-2(M-\omega_{2})^{2}+(M-m_{2})\sqrt{4(M-m_{2})^{2}+2a^{2}}-(M-\omega_{2})\sqrt{4(M-\omega_{2})^{2}+2a^{2}}
\nonumber\\-2(M-(m_{1}+m_{2}))^{2}+2(M-(\omega_{1}+\omega_{2}))^{2}-(M-(m_{1}+m_{2}))\sqrt{4(M-(m_{1}+m_{2}))^{2}+2a^{2}}
\nonumber\\+(M-(\omega_{1}+\omega_{2}))\sqrt{4(M-(\omega_{1}+\omega_{2}))^{2}+2a^{2}}\bigg\},
\end{eqnarray}
which again is not zero. Again, with respect to the correlation function $$\chi=8\pi\big(\omega_{1}\omega_{2}-m_{1}m_{2}\big)$$ for the ordinary massive case, in this study the role of the quantum deformation is to amplify the correlation function between radiating massive particles.

\section{Conclusions}

It is well known that taking gravitational effects into account makes quantum mechanics and quantum field theory regular in the ultraviolet sector. But on the other hand, bringing quantum field theory results into outcomes and phenomena arising from the general theory of relativity makes these subjects theoretically deform to answer some fundamental problems of physics, including black hole radiation, singularities, the information loss problem, etc. In this paper we have generalized the famous Parikh–Wilczek tunneling mechanism to a more viable framework where quantum effects are taken into account. We have used the quantum-deformed Schwarzschild line element suggested by Kazakov and Solodukhin. The main outcomes of this study are as follows:

\begin{itemize}
\item{Quantum deformation of the Schwarzschild line element as suggested by Kazakov and Solodukhin results in a non-zero-mass black hole remnant with finite temperature at the final stage of evaporation. This remnant could be essentially a Planck-scale remnant, and could also be a potential candidate for dark matter.}
\item{A quantum-deformed Schwarzschild black hole mimics the behavior of a charged classical Reissner-Nordstr\"{o}m black hole. This similarity in our framework is provided by the interesting relation $Q\equiv i\frac{\sqrt{2}}{2}a$, where $Q$ is the electric charge parameter of the Reissner-Nordstr\"{o}m black hole. This extraordinary resemblance tempts us to argue that there is some connection (and even a close relation) between electric charge and quantum effects. In other words, electric charge may originate from some as yet unknown quantum effects. The nature of the parameter $a$ in the Kazakov and Solodukhin solution may finally be a clue towards understanding this amazing feature. This resemblance has been observed in a non-commutative framework too~\cite{Nozari2008c}}.
\item{The Hawking radiation in this quantum-deformed tunneling scenario is not completely thermal. Indeed, the role of quantum deformation is to strengthen the non-thermality of Hawking radiation. There are correlations between emitted particles (modes) so that part of the information may be attributed to these correlations. This feature possibly sheds light on the information loss problem.}
\end{itemize}

{\bf Acknowledgement}\\
The authors would like to thank Sara Saghafi, Sareh Eslamzadeh, and Narges Rashidi for fruitful discussions.

\end{document}